\documentclass[useAMS,usenatbib]{mnras}
\usepackage{array}
\usepackage{times}
\usepackage{graphicx}
\usepackage{mathptmx}
\usepackage{amsmath}
\usepackage{amssymb}
\usepackage[usenames]{color}
\def\mnras{MNRAS}
\def\apjl{ApJL }
\def\aj{AJ }
\def\apj{ApJ }

\def\pasp{PASP }

\def\apjs{ApJS }
\def\araa{ARAA }
\def\aap{A\&A }
\def\nat{Nature }
\usepackage{hyperref}
\hypersetup{colorlinks=true,citecolor=blue,linkcolor=purple,filecolor=black,runcolor=black}
\begin{document}
\newcommand{\comment}[1]{}
\definecolor{purple}{RGB}{160,32,240}
\newcommand{\peter}[1]{\textcolor{purple}{(\bf #1)}}
\newcommand{\macc}{M_\mathrm{acc}}
\newcommand{\mpeak}{M_\mathrm{peak}}
\newcommand{\mnow}{M_\mathrm{now}}
\newcommand{\vacc}{v_\mathrm{acc}} 
\newcommand{\vpeak}{v_\mathrm{peak}} 
\newcommand{\vnow}{v^\mathrm{now}_\mathrm{max}}

\newcommand{\Mnfw}{M_\mathrm{NFW}}
\newcommand{\Msun}{\;\mathrm{M}_{\odot}}
\newcommand{\mvir}{M_\mathrm{vir}}
\newcommand{\rvir}{R_\mathrm{vir}}
\newcommand{\vmax}{v_\mathrm{max}}
\newcommand{\vmac}{v_\mathrm{max}^\mathrm{acc}}
\newcommand{\mvac}{M_\mathrm{vir}^\mathrm{acc}}
\newcommand{\sfr}{\mathrm{SFR}}
\newcommand{\plotgrace}[1]{\includegraphics[width=\columnwidth,type=pdf,ext=.pdf,read=.pdf]{#1}}
\newcommand{\plotssgrace}[1]{\includegraphics[width=0.95\columnwidth,type=pdf,ext=.pdf,read=.pdf]{#1}}
\newcommand{\plotgraceflip}[1]{\includegraphics[width=\columnwidth,type=pdf,ext=.pdf,read=.pdf]{#1}}
\newcommand{\plotlargegrace}[1]{\includegraphics[width=2\columnwidth,type=pdf,ext=.pdf,read=.pdf]{#1}}
\newcommand{\plotlargegraceflip}[1]{\includegraphics[width=2\columnwidth,type=pdf,ext=.pdf,read=.pdf]{#1}}
\newcommand{\plotminigrace}[1]{\includegraphics[width=0.5\columnwidth,type=pdf,ext=.pdf,read=.pdf]{#1}}
\newcommand{\plotmicrograce}[1]{\includegraphics[width=0.25\columnwidth,type=pdf,ext=.pdf,read=.pdf]{#1}}
\newcommand{\plotsmallgrace}[1]{\includegraphics[width=0.66\columnwidth,type=pdf,ext=.pdf,read=.pdf]{#1}}
\newcommand{\plotappsmallgrace}[1]{\includegraphics[width=0.33\columnwidth,type=pdf,ext=.pdf,read=.pdf]{#1}}

\newcommand{\hinv}{h^{-1}}
\newcommand{\mpc}{\rm{Mpc}}
\newcommand{\hmpc}{$\hinv\mpc$}

\title{On The History and Future of Cosmic Planet Formation}

\author[Behroozi \& Peeples]{Peter~S.~Behroozi$^{1}$,\thanks{E-mail:     behroozi@stsci.edu} Molly Peeples$^{1}$
\\
\\
$^{1}$ Space Telescope Science Institute, Baltimore, MD 21218, USA\\
}
\date{Released \today}
\pubyear{2015}

\maketitle

\begin{abstract}
We combine constraints on galaxy formation histories with planet formation models, yielding the Earth-like and giant planet formation histories of the Milky Way and the Universe as a whole.  In the Hubble Volume (10$^{13}$ Mpc$^3$), we expect there to be $\sim 10^{20}$ Earth-like and $\sim 10^{20}$ giant planets; our own galaxy is expected to host $\sim10^{9}$ and $\sim10^{10}$ Earth-like and giant planets, respectively.  Proposed metallicity thresholds for planet formation do not significantly affect these numbers.  However, the metallicity dependence for giant planets results in later typical formation times and larger host galaxies than for Earth-like planets.  The Solar System formed at the median age for existing giant planets in the Milky Way, and consistent with past estimates, formed after 80\% of Earth-like planets.  However, if existing gas within virialised dark matter haloes continues to collapse and form stars and planets, the Universe will form over 10 times more planets than currently exist.  We show that this would imply at least a 92\% chance that we are not the only civilisation the Universe will ever have, independent of arguments involving the Drake Equation.
 \end{abstract}
\begin{keywords}
planets and satellites: terrestrial planets, planets and satellites: gaseous planets, galaxies: formation
  \end{keywords}

\section{Introduction}

\label{s:introduction}

Early estimates of the planet formation history of the Universe \citep{Livio99,Lineweaver01} suggested that the Earth formed after 75--80\% of other similar planets, even when considering potential galactic habitable zones \citep{Lineweaver04}.  Since that time,  thousands of exoplanets have been found, aided by the \textit{Kepler} mission \citep{Lissauer14}.  Many advances have been made in the past decade, especially in our understanding of how planet formation depends on the mass and metallicity of the host star \citep{Fischer05,Buchhave12,Wang13Planet,Lissauer14,Buchhave14,Gonzalez14Planet,Reffert14}.   Concurrently, constraints on galaxies' star formation and metallicity histories have been improving rapidly \citep{Maiolino08,Mannucci10,Moustakas11,BWC13,Peeples13,Munoz14}.

In this paper, we combine recent planet frequency models \citep{Lissauer14,Buchhave14} with reconstructed galaxy formation histories \citep{Maiolino08, BWC13} to update constraints on the planet formation history of the Milky Way and the Universe as a whole, both for Earth-like planets and for giant planets.  We adopt a flat, $\Lambda$CDM cosmology with $\Omega_M = 0.27$, $\Omega_\Lambda = 0.73$, $h=0.7$, $n_s = 0.95$, and $\sigma_8 = 0.82$, similar to recent WMAP9 constraints \citep{WMAP9}; the initial mass function (IMF) is assumed to follow \cite{Chabrier03} from 0.1 to 100 $\Msun$.

\newcommand{\repos}[3]{\hspace{-#1}\raisebox{#2}{\llap{#3}}\hspace{#1}}

\begin{figure*}
\vspace{-3ex}
\begin{tabular}{cc}
\plotgrace{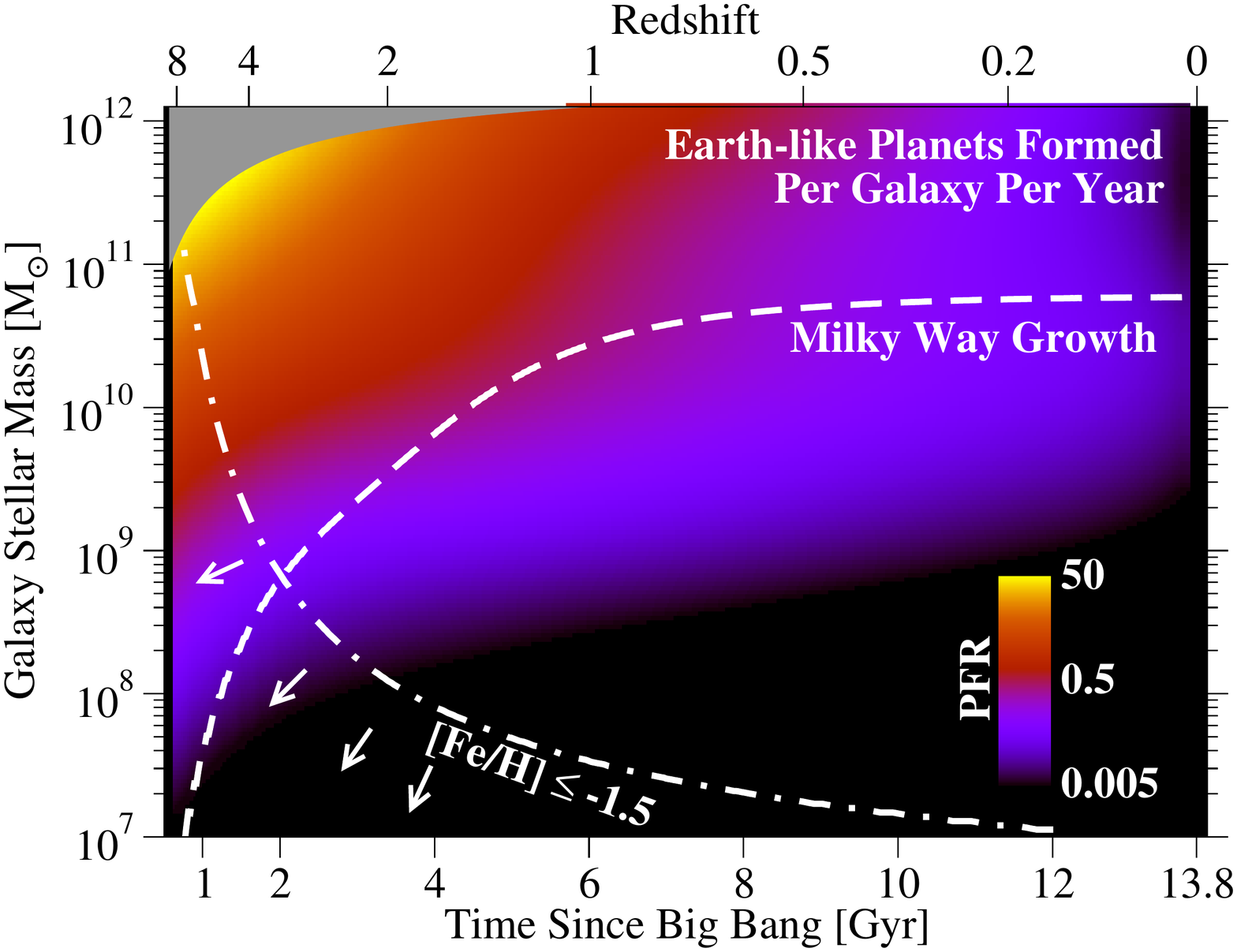} & \plotgrace{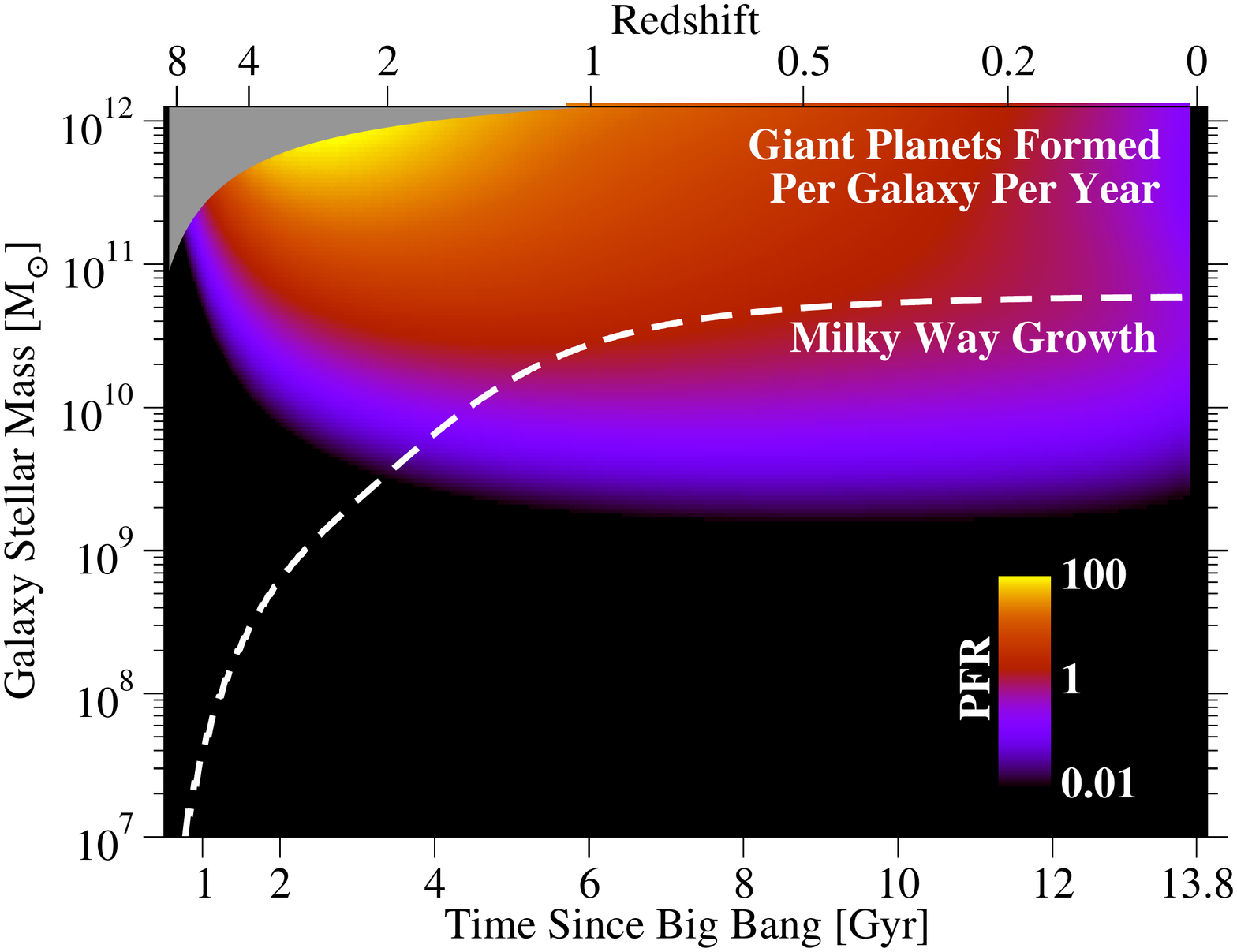}\\[-3ex]
\plotgrace{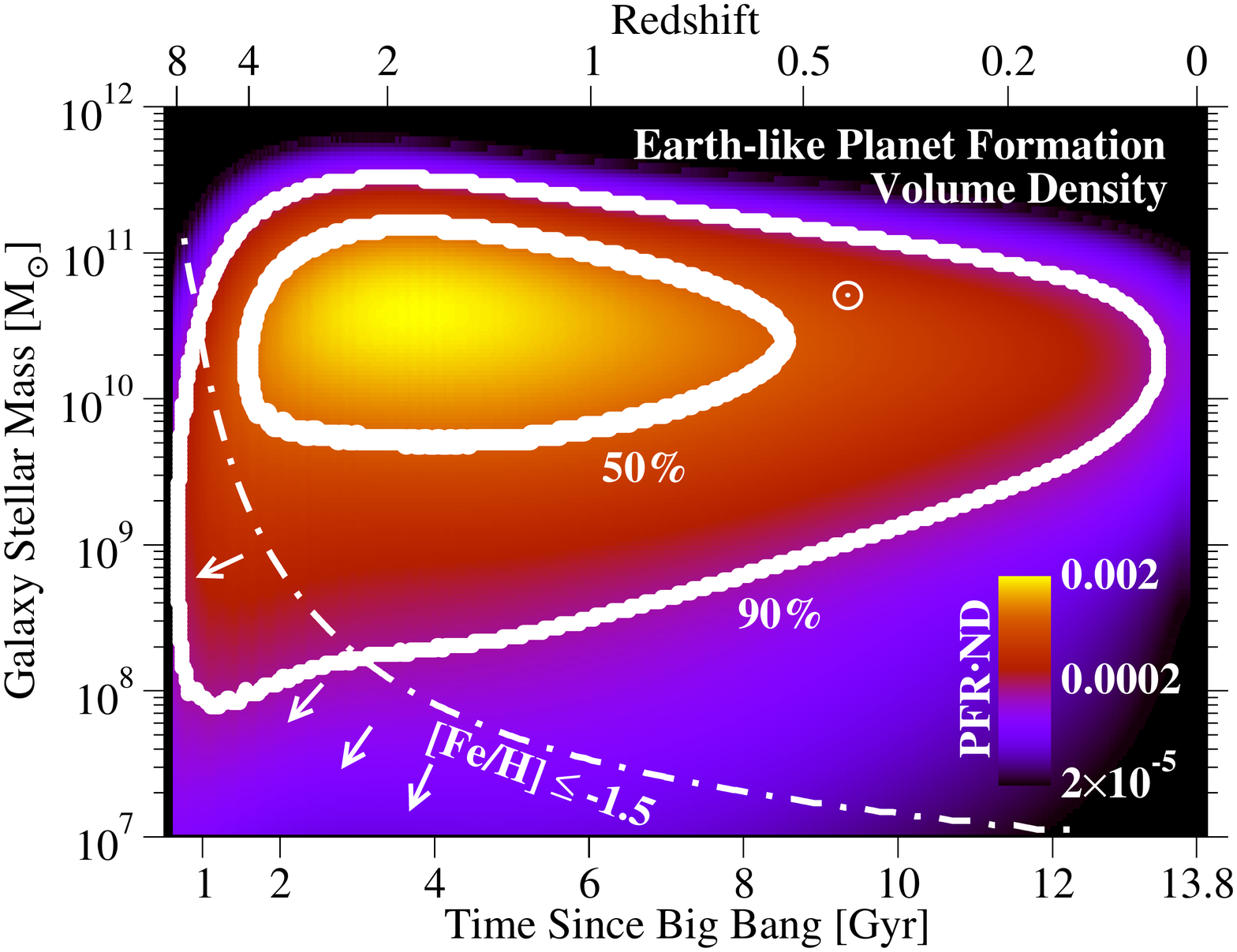} & \plotgrace{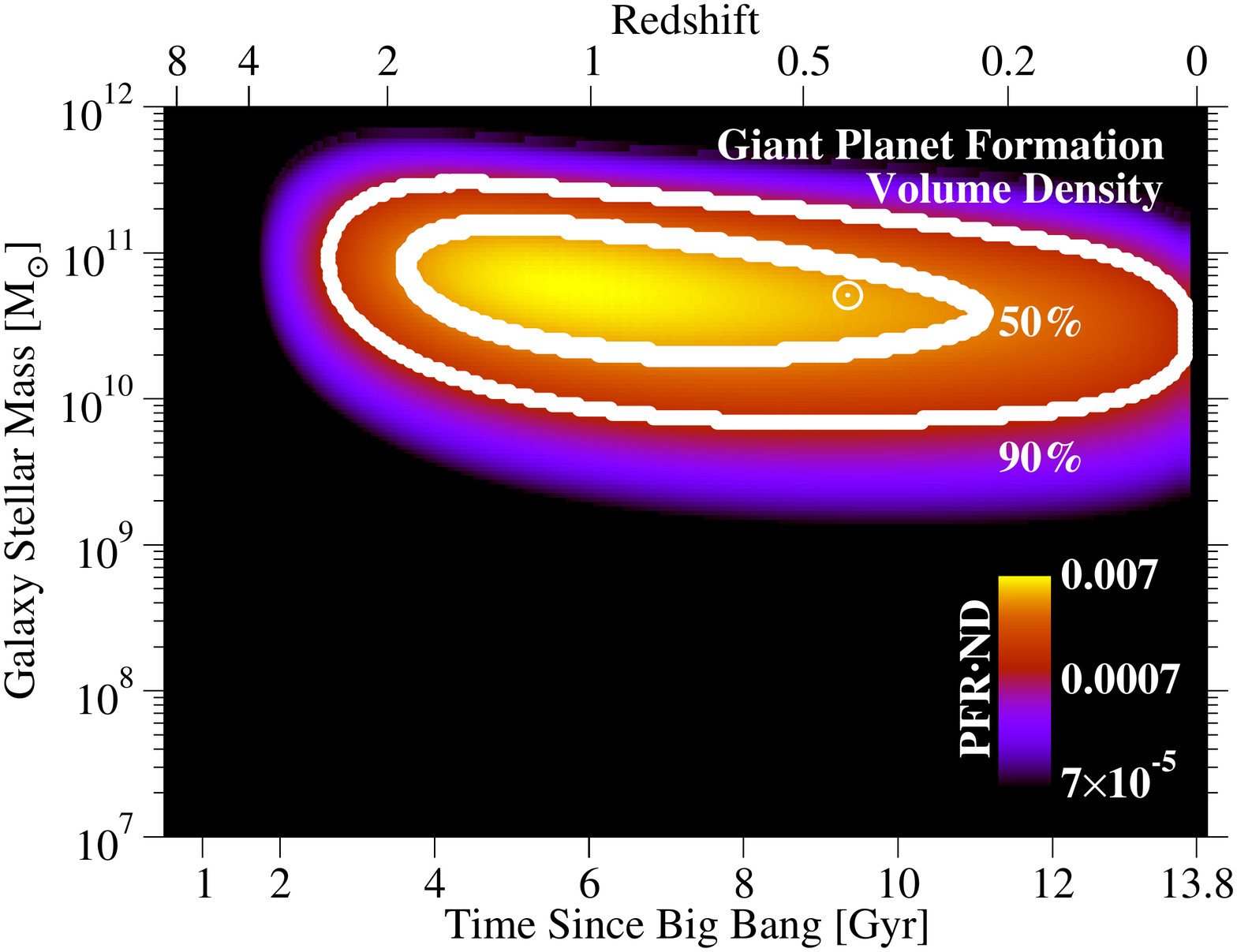} \\
\end{tabular}\\
\caption{\textbf{Top-left} panel: formation rate (in planets/yr) for Earth-like planets as a function of galaxy stellar mass and cosmic time.  The dashed line indicates the median expected growth history of the Milky Way \citep{BWC13}.  The dot-dashed line indicates [Fe/H]=$-1.5$, which has been suggested \citep{Johnson12} as the threshold metallicity for planet formation.  Grey shaded areas indicate where galaxies are not expected to exist in the observable Universe.  \textbf{Top-right} panel: same, for giant planets.  \textbf{Bottom-left} panel: Earth-like planet formation rate multiplied by galaxy number density as a function of stellar mass and cosmic time, i.e., the volume density of planet formation (in planets/yr/comoving Mpc$^3$/dex).  Contours indicate where 50\% and 90\% of all planet formation has taken place.  The $\odot$ symbol indicates the Milky Way's stellar mass and age at the formation of the Solar System.  \textbf{Bottom-right} panel: same, for giant planets.}
\label{f:model}
\end{figure*}

\begin{figure*}
\vspace{-3ex}
\begin{tabular}{cc}
\plotgrace{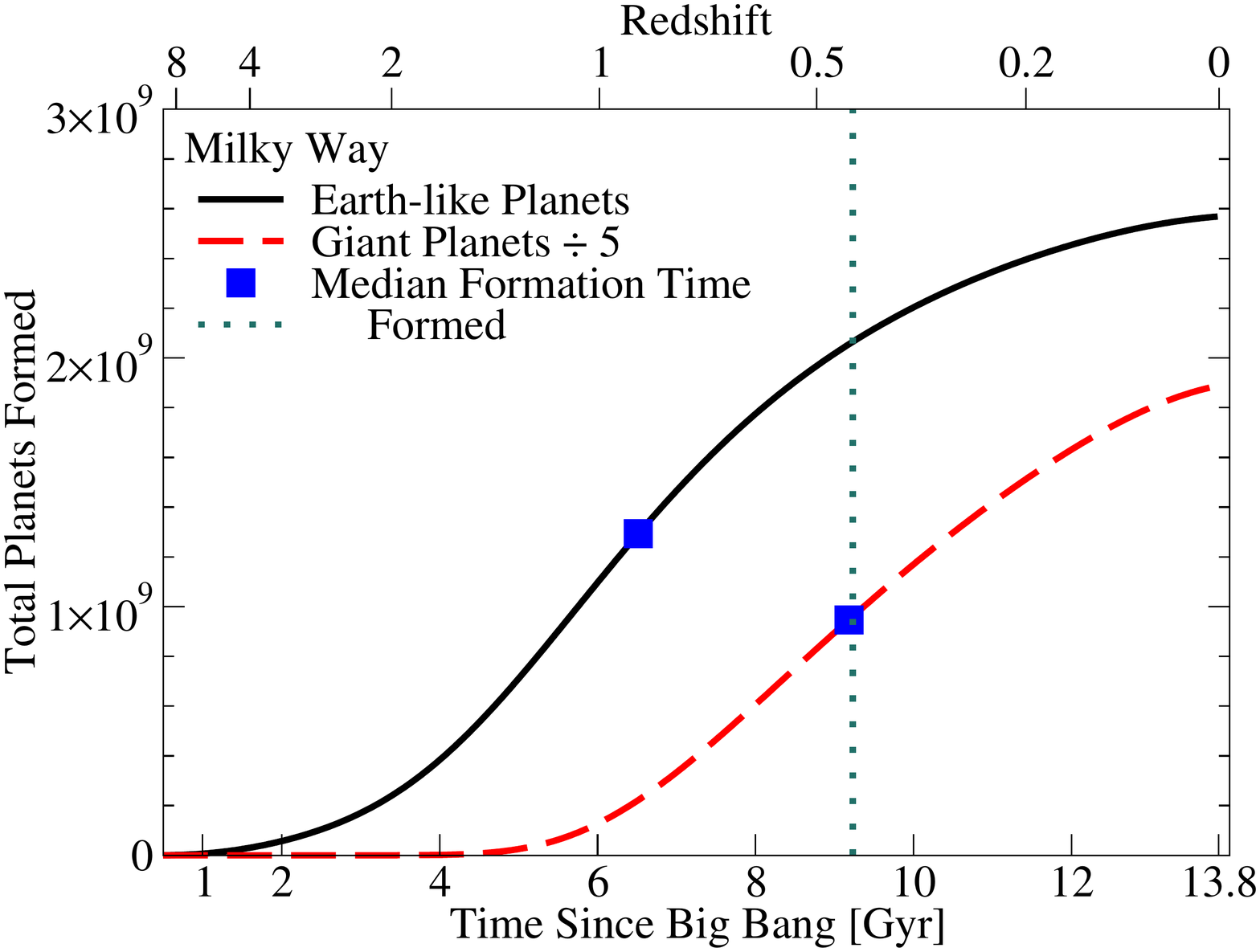}\repos{44.7ex}{32.8ex}{$\boldsymbol\odot$} & \plotgrace{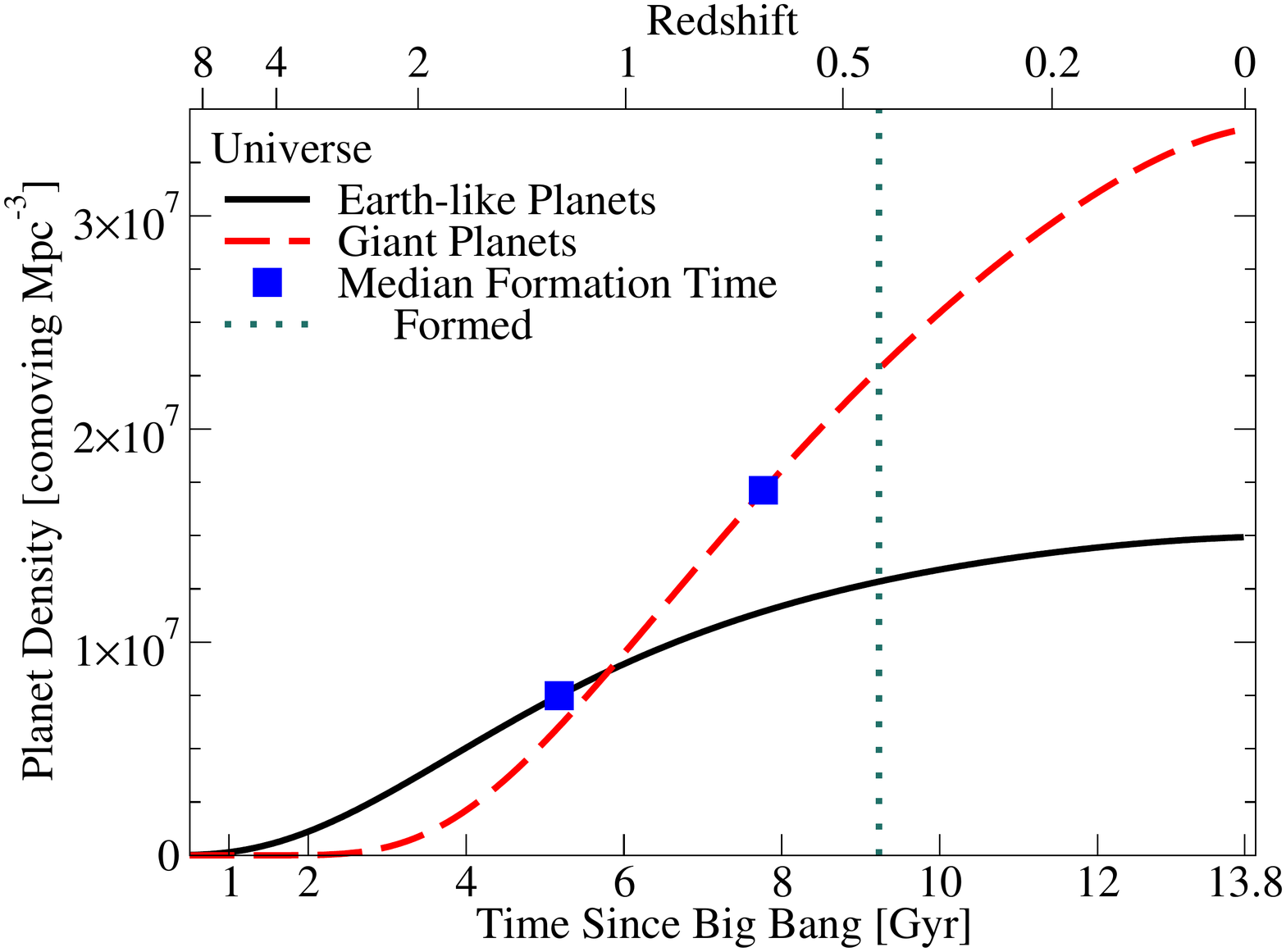}\repos{44.7ex}{32.8ex}{$\boldsymbol\odot$}
\end{tabular}\\[-5ex]
\begin{tabular}{cc}
\plotgrace{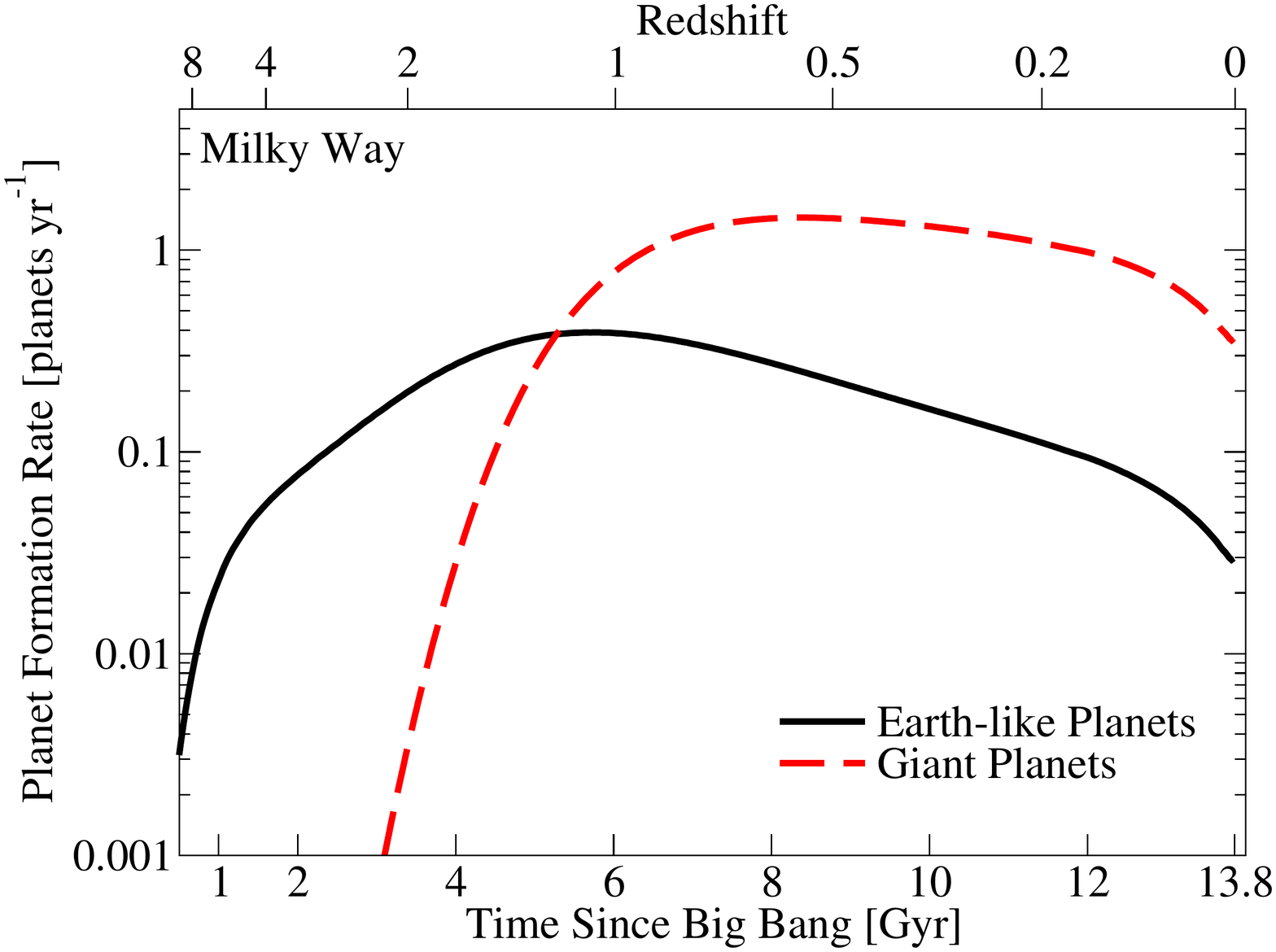}& \plotgrace{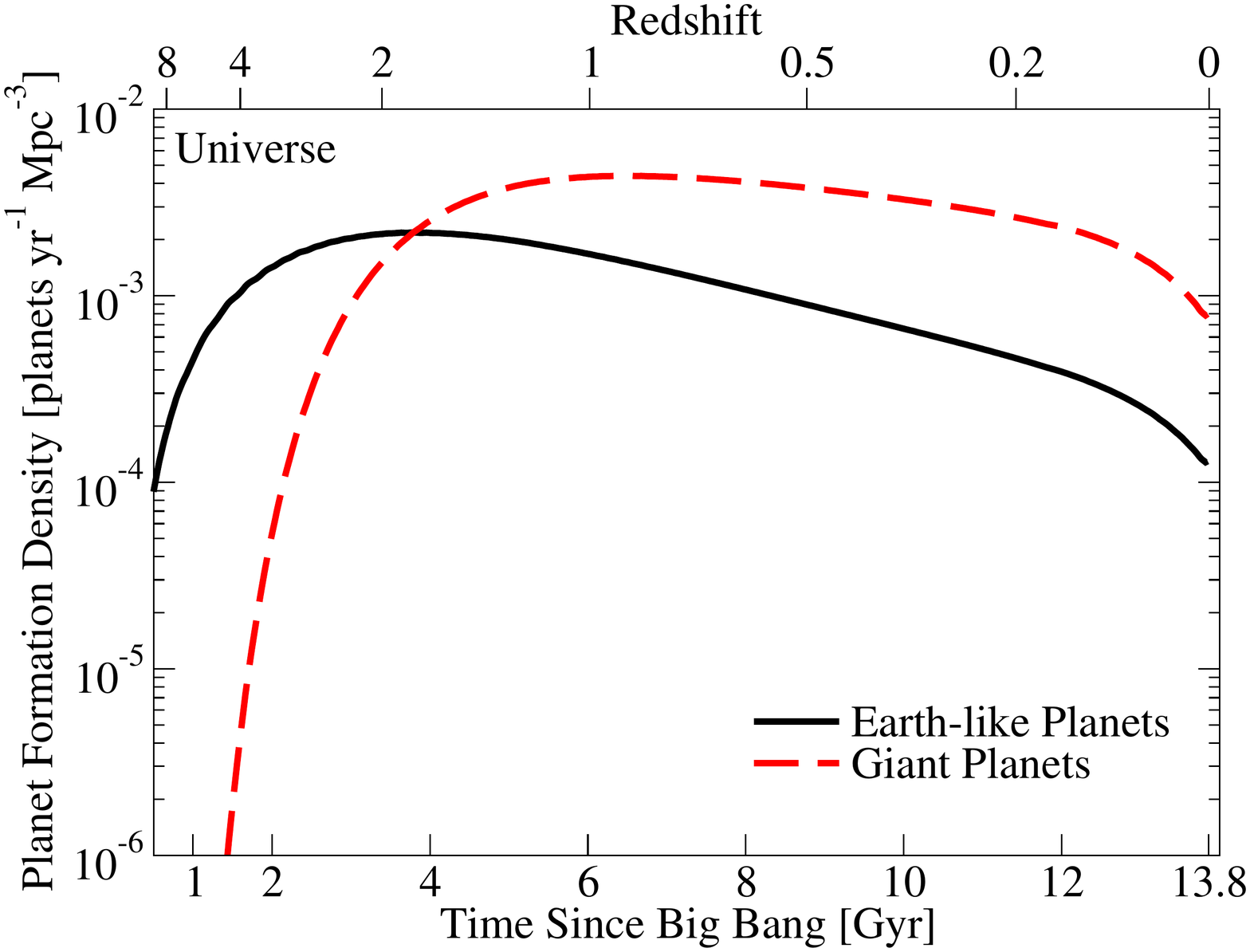}
\end{tabular}
\caption{\textbf{Top-left} panel: total Earth-like and giant planets formed in the Milky Way as a function of cosmic time.  Giant planet counts have been shifted by a factor of 5 to allow better comparison with the Earth-like planet formation history.  \textbf{Top-right} panel: average planet density in the Universe as a function of cosmic time.  Earth-like planet formation tracks the galaxy/cosmic star formation rates, whereas giant planet formation times are greater at late times due to their metallicity dependence.  Blue squares mark the median formation times of each population.  The vertical dotted line indicates the formation time of the Solar System, which occurred after 80\% of present-day Earth-like planets and 50\% of present-day giant planets were formed in the Milky Way. \textbf{Bottom} panels: planet formation rates and densities, respectively, for the Milky Way and the Universe as a whole.  Uncertainties in all estimates are $\sim 1$ dex, arising from uncertainties in planet detection rates with Kepler.}
\label{f:model2}
\end{figure*}

\section{Methodology}

\label{s:methodology}

While planet formation can depend on host star mass, new stars' masses are drawn from a nearly universal distribution \citep{Chabrier03}.  When averaged over an entire galaxy, the planet formation rate (PFR) is then proportional to the galaxy's star formation rate, modified by the PFR's metallicity dependence.  Using a power-law parametrisation for this metallicity dependence \citep[see, e.g.,][]{Fischer05,Gonzalez14Planet}, we model the planet formation rate of a galaxy as a function of its stellar mass ($M_*(t)$) and cosmic time ($t$):
\begin{equation}
\label{e:pfr}
\mathrm{PFR}(M_*,t) = n \frac{(Z(M_\ast, t)/Z_\odot)^\alpha}{\langle m_\ast \rangle} \mathrm{SFR}(M_\ast, t),
\end{equation}
where $n$ is the mean number of planets formed per star, $\alpha$ is the power-law dependence of planet incidence on metallicity, $Z(M_*, t)$ is the galaxy's mean gas-phase metallicity, $\langle m_\ast \rangle = 0.67\Msun$ is the mean mass of a newly-formed star \citep{Chabrier03}, and $\mathrm{SFR}(M_\ast, t)$ is the galaxy's star formation rate in $\Msun$ yr$^{-1}$.  Additional factors (e.g., stellar initial mass functions and densities) influencing the PFR are discussed in Appendix \ref{a:systematics}.

For giant planets ($R > 6\;\mathrm{R}_\oplus$; including, e.g., Jupiter and Saturn), the metallicity (specifically, [Fe/H]) dependence is long-established \citep{Fischer05}; recent estimates suggest $n_G \sim 0.022$ and $\alpha_G \sim 3.0$ \citep{Gonzalez14Planet}, albeit with significant systematic uncertainties.  To define Earth-like (i.e., ``habitable zone'') planets, we adopt the same definition as \cite{Lissauer14}, requiring that planets with an Earth-like atmosphere could support stable surface reservoirs of liquid water.  Effectively, this includes all objects whose radii and orbital periods are within a factor of $e$ of those of the Earth \citep{Lissauer14}.  In the Solar System, this would include Mars and Venus, but exclude, e.g., Mercury and the Moon.  The metallicity dependence for Earth-like planets is believed to be smaller than for giant planets \citep{Buchhave12,Campante15}, with recent estimates suggesting $\alpha_E \sim 0$ to $0.7$ \citep{Wang13Planet,Lissauer14}.  This range of $\alpha_E$ has only a small impact on our results, so we conservatively take $\alpha_E = 0$ (i.e., no metallicity dependence) for Earth-like planets.  However, \cite{Johnson12} suggest a theoretical minimum metallicity threshold for Earth-like planet formation of [Fe/H]$\sim-1.5 + \log_{10}\left(\frac{r}{\mathrm{AU}}\right)$ (with $r$ the orbital radius), so we mark a fiducial threshold of [Fe/H]$=-1.5$ in all relevant plots.  For $n_E$, Kepler has provided the largest statistical samples \citep{Catanzarite11}; \cite{Lissauer14} suggest an incidence of $\sim 0.1$ Earth-like planets per Sun-like star.  Habitable zones are expected to exist only around 0.6-1.4 $\Msun$ ($K$ to $F5$-class) stars \citep{Kasting93,Kopparapu14}, which make up 14.8\% of stars by number \citep{Chabrier03}, so we take $n_E = 0.015$.

\cite{BWC13} determined $\mathrm{SFR}(M_\ast, t)$ for galaxies up to $\sim$ 13 Gyr ago ($z=8$), covering ${>90\%}$ of all star formation \citep{BehrooziEvolution}.  The methodology is detailed in Appendix \ref{a:bwc13}; briefly, it involves linking galaxies at one redshift to galaxies with the same cumulative number densities at another redshift to trace their stellar mass buildup, as the most massive galaxies at one redshift will tend to remain the most massive galaxies at later redshifts.  The full computation also involves corrections for scatter in galaxy growth histories and galaxy-galaxy mergers \citep{BehrooziND,BehrooziTree}.  Knowing the stellar mass history of galaxies, one may use observed metallicity--stellar mass--redshift relations \citep[e.g.,][]{Maiolino08,Moustakas11} or metallicity--stellar mass--star formation rate relations \citep[e.g.,][]{Mannucci10} to determine galaxy metallicity histories \citep[see also][]{Munoz14}.  Here, we use the fitting function in \cite{Maiolino08}, which is constrained for $z<3.5$ ($<$11.7 Gyr ago), mildly extrapolated over the same redshift range as our star formation histories.  As \cite{Maiolino08} measure oxygen abundance ratios ([O/H]), we use the formula [Fe/H]$ = -0.1 + 1.182$ [O/H] (from fitting Milky Way stellar abundances in the SDSS-III APOGEE; \citealt{Holtzmann15}) to convert to iron abundance ratios.  If all stars instead had solar iron-to-oxygen ratios, our derived giant planet abundance would increase by $\sim 0.3$ dex.

\section{Results}

\label{s:results}

The resulting planet formation rates from Eq.\ \ref{e:pfr} are shown in the top panels of Fig.\ \ref{f:model}.  Similar to star formation \citep{BWC13}, planet formation rates per galaxy are greatest at early times in massive galaxies.  Indeed, Earth-like planet formation rates are exactly proportional to galaxy star formation rates (scaled by $n_E=0.0073$) in our assumed model.  The planet formation history of the Milky Way (present-day $M_\ast = 5-7 \times 10^{10}\Msun$, from current Bayesian models; \citealt{Licquia14}) can be inferred by integrating PFR$(M_*, t)$ along the median growth history for Milky Way-sized galaxies (dashed line, top panels of Fig.\ \ref{f:model}).  The planet formation history of the Universe, here expressed as the average volume density of planet formation, is obtained by multiplying PFR$(M_*, t)$ by the volume density of galaxies as a function of stellar mass and cosmic time ($\phi(M_*, t)$, discussed in Appendix \ref{a:bwc13}).

The product of PFR with $\phi$ shows when and where all Earth-like and giant planets formed (Fig.\ \ref{f:model}, bottom panels).  Typical galaxy masses at the time of planet formation are $\sim 10^{10.5}\Msun$, regardless of planet type.  However, because of their metallicity bias, giant planets form later than Earth-like planets.  Giant planet formation is rare in galaxies below 10$^{9}\Msun$; while the Magellanic Clouds may have some giant planets, it is unlikely that lower-mass dwarf satellite galaxies of the Milky Way will have any.  In both cases, the \cite{Johnson12} minimum metallicity threshold is a weak one, as the vast majority of star formation has taken place at [Fe/H]$>-1.5$.  We find that total planet densities would be lowered by $<10\%$ for Earth-like planets and $\ll 0.01\%$ for giant planets with this metallicity threshold (Fig.\ \ref{f:model}, bottom panels).  However, this threshold would strongly diminish the number of Earth-like planets formed around stars older than 12 billion years in the Milky Way (Fig.\ \ref{f:model}), which is so far consistent with the age of the oldest observed star with Earth-like companions ($11.2\pm 1.0$ Gyr; \citealt{Campante15}). 

We show total planet formation histories and rates from Eq.\ \ref{e:pfr} for the Milky Way and the Universe in Fig.\ \ref{f:model2}.  In the Universe's observable volume ($10^{13}$ Mpc$^3$), these results would imply $\sim 10^{20}$ Earth-like planets and a similar number of giant planets.\footnote{The number of \textit{observable} planets in our past lightcone, as opposed to planets currently in the Hubble volume, is $\sim 10^{19}$ for both Earth-like and giant planets.}  Errors are dominated by planet incidence rates ($n_E$ and $n_G$), which are uncertain at the 0.5-1 dex level \citep{Lissauer14} due to different detection efficiency estimates.  Smaller systematic errors (0.2-0.3 dex) come from uncertainties in initial mass functions, stellar population modelling in galaxies, and variation in individual galaxy star formation histories and metallicities \citep{Behroozi10,Peeples14}.

\section{Discussion}

We discuss the Solar System's relative formation time (\S \ref{s:form}) and its relation to the expected number of future civilisations (\S \ref{s:prob}).

\subsection{Formation Time of the Solar System}

\label{s:form}

\begin{figure}
\plotgrace{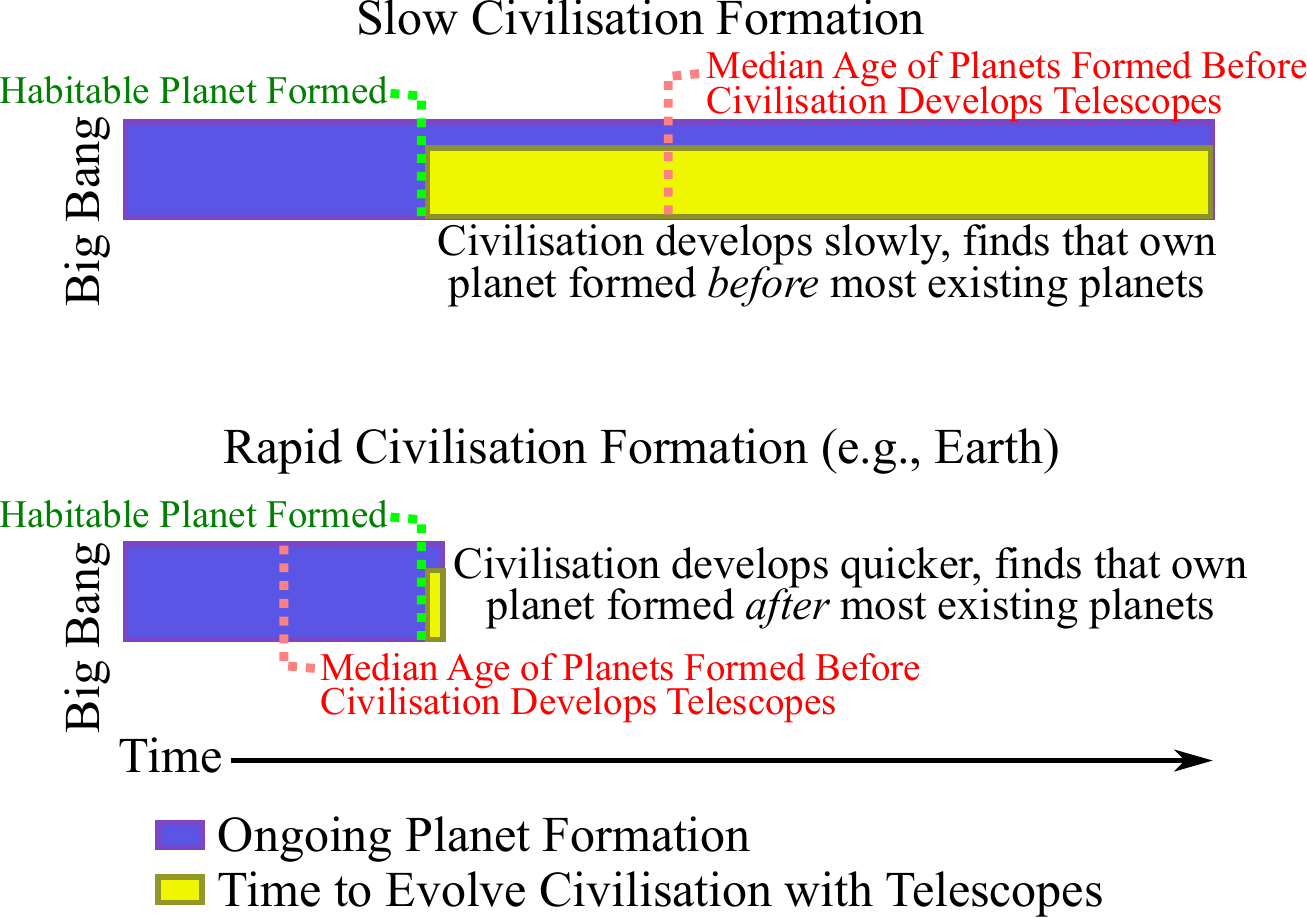}
\caption{The relative formation time of one's \textit{own} planet depends on the time it takes one's civilisation to form.  As shown above, planet formation continues while civilisations are developing.  Many planets will form if a civilisation is slow to develop, so by the time it is able to calculate its own planet's formation time relative to others ($\sim$ the time when it develops telescopes), it will find that its planet formed early.  In contrast, a rapidly-developing civilisation (e.g., ours) reaches that stage earlier, giving the Universe less time to make more planets; the civilisation will then find that its own planet formed late relative to most others.}
\label{f:civ_time}
\vspace{-3ex}
\plotgrace{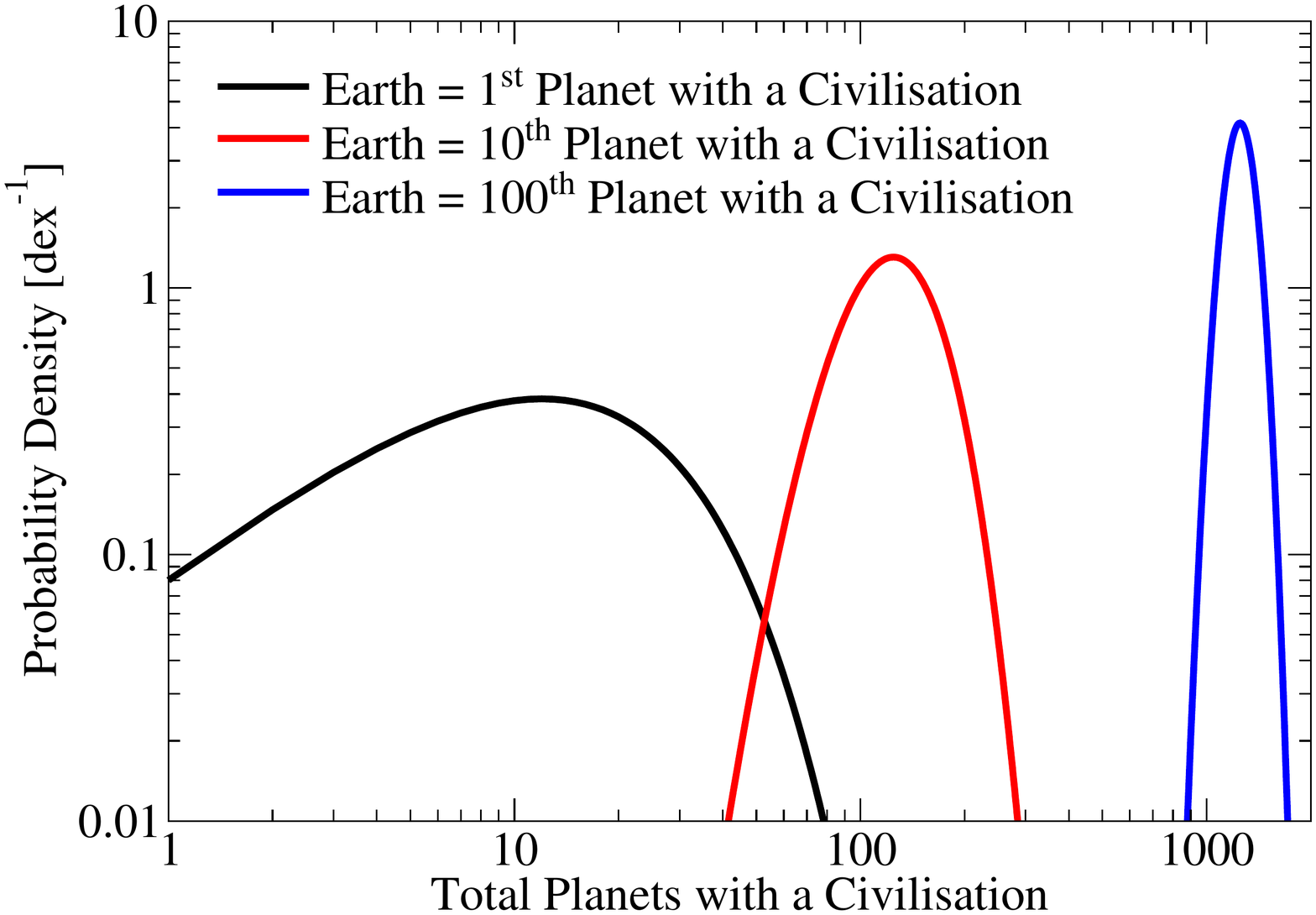}
\caption{Probability for the total number of planets with civilisations in the Universe, given that the Earth formed before 92\% of similar planets expected to exist.  If Earth is the $n$th planet since the Big Bang to have formed a civilisation, then the average (expected) total number of planets with civilisations scales as $12.5n$.  Even for the most conservative possible assumption (i.e., that Earth was the 1$^\mathrm{st}$ planet formed that evolved an intelligent civilisation), it is unlikely that we will be the only civilisation that the Universe will ever have (\textit{black line}).  As the number of earlier planets with civilisations increases (\textit{red and blue lines}), it becomes more and more likely that the Universe will have many more civilisations than currently exist.  For comparison, if the Milky Way today contained another civilisation, it is likely that Earth would be at least the ten billionth planet to host a civilisation in the observable universe, which would eventually contain at least a  hundred billion civilisations.}
\label{f:civs}
\end{figure}

Fig.\ \ref{f:model2} shows that the Earth formed later than $\sim 80\%$ of similar planets in both the Milky Way and the Universe, matching previous findings \citep{Livio99,Lineweaver01}. Comparatively, the Solar System (including Jupiter) formed closer to the median formation time for giant planets.  This is not evidence for or against giant planets being prerequisite for life as there is a strong observer bias (Fig.\ \ref{f:civ_time}).  When calculating the age of our \textit{own} planet, we are really calculating the time $t_c$ that it took our own species and civilisation to evolve.   If $t_c$ were extremely long, many new planets would have formed later than our own planet but before intelligent life evolved---so we would have concluded that our planet formed early compared to most other planets.  However, as $t_c$ is shorter than the current doubling time, $t_d$, for stellar mass in the Milky Way ($t_c = 4.6$ Gyr and $t_d \sim 20$ Gyr), fewer planets have had time to form while civilisation has developed.  Hence, the ``late'' formation time of our own planet speaks more to the ratio of $t_c$ to $t_d$ than to conditions for habitability.

This observer bias can be removed if we calculate our formation time relative to all the planets which will ever be formed.  The Milky Way is expected to merge with Andromeda (M31) in $\sim$4 Gyr \citep{Cox08}, forming a single object with total (dark matter and baryonic) mass 3.17$\times 10^{12}\Msun$  \citep{Marel12}.  Using fitting formulae in \cite{BehrooziUnbound} for its continued mass growth, we expect that its total mass will asymptote to 3.9$\times 10^{12}\Msun$.  Haloes of these masses are expected to have approximately the cosmic baryon fraction (16.5\%; \citealt{WMAP9}) of their mass in gas and stars \citep{Werk14}, which translates to 6.4$\times 10^{11}\Msun$ of baryonic matter within the eventual halo.  The present-day combined stellar masses of the Milky Way \citep{Licquia14} and M31 \citep{Tamm12} are $\sim1.8\times 10^{11}\Msun$; correcting for the $\sim 30\%$ of stellar mass lost in normal stellar evolution \citep{Chabrier03}, this leaves 3.9$\times 10^{11}\Msun$ of gas in the halo available for future star formation.  As the remaining gas eventually cools and forms stars (as is expected to occur over the next trillion years; \citealt{Adams97,Tutukov00,Nagamine04}), this implies that the Earth has actually formed earlier than $\sim$ 61\% of all planets that will ever form in the Milky Way--M31 group.

Repeating this calculation for the Universe as a whole, we note that only 8\% of the currently available gas around galaxies (i.e., within dark matter haloes) had been converted into stars at the Earth's formation time \citep{BehrooziEvolution}.  Even discounting any future gas accretion onto haloes, continued cooling of the existing gas would result in Earth having formed earlier than at least 92\% of other similar planets.  For giant planets, which are more frequent around more metal-rich stars, we note that galaxy metallicities rise with both increasing cosmic time and stellar mass \citep{Maiolino08}, so that future galaxies' star formation will always take place at higher metallicities than past galaxies' star formation.  As a result, Jupiter would also have formed earlier than at least $\sim 90\%$ of all past and future giant planets.

As shown in Fig. \ref{f:model2}, planet formation rates have declined significantly since $z\sim 2$ (for Earth-like planets) and $z\sim 1$ (for giant planets), primarily because of declines in the cosmic star formation rate.  If these declines continue, most of the additional planets formed in both the Universe and the Milky Way will be in the very far future (100 Gyr to 1 Tyr from now) compared to the current age of the Universe ($\sim 13.8$ Gyr; \citealt{WMAP9}).  Hence, as the Universe's accelerating expansion is rapidly reducing the number of observable galaxies \citep{Loeb02,Nagamine03}, most future planets formed in other galaxies will not be visible from the Milky Way.

\subsection{Probability of Other Civilisations}

\label{s:prob}

The Drake equation \citep{Drake92} for calculating the number of intelligent, communicative civilisations (hereafter, just ``civilisations'') is famously uncertain, with estimates of the civilisation incidence per habitable planet ranging from $10^{-5}$ \citep{Sagan63} to arbitrarily small values (e.g., $<10^{-30}$, combining pessimistic estimates from \citealt{Carter83,Ward00,Schermer02,Spiegel12}).  Combined with our estimates of the number of Earth-like planets (\S \ref{s:results}) and the fact of our existence, this would result in 1 to $10^{15}$ civilisations in the Universe and 1 to $10^{4}$ in the Milky Way at the present time.

The formation time of our planet (compared to all which will ever form) gives weak but independent constraints on the total number of planets with civilisations which will ever exist.  Intuitively, if we were the only civilisation the Universe will ever have, the Copernican principle suggests that it is unlikely for our planet to have formed so early relative to other similar ones.\footnote{Our formation time relative to other habitable planets in the Milky Way is unexceptional (\S \ref{s:form}), giving little information on the total number of civilisations our galaxy will have.  Additional systematics for how planet migration/scattering affect civilisation formation are discussed in Appendix \ref{a:systematics}.}    As an example, we can calculate an upper bound for the chance that the Universe will only ever have a single civilisation (corresponding to civilisation incidences of $<10^{-21}$ per habitable planet).  For this upper bound, we adopt the prior that Earth is the first planet with a civilisation to have formed---any possibility that Earth is \textit{not} the first is incompatible with there being only one civilisation.  We find using Bayes' rule that the probability of there being $N$ civilisations total is then:
\begin{equation}
P(N|f,E=1) \propto N f^{N-1} P(N)
\label{e:bayes}
\end{equation}
where $f=0.92$ is the fraction of planets which have yet to form, $E=1$ is the assumption that the Earth has the first civilisation, and $P(N)$ is the prior on the number of planets with civilisations.  Because Eq.\ \ref{e:bayes} falls off exponentially for large civilisation numbers $N$, it is only necessary to know the prior $P(N)$ over a modest dynamic range ($1\le N \le 1000$).  The orders-of-magnitude uncertainties on parameters in the Drake Equation suggest that the prior $P(N)$ on the number of planets with civilisations should be a log-normal distribution with an exceptionally large width ($\gtrsim 20$ dex).  Locally, then, it is an excellent approximation to take $P(N)$ as uniform in logarithmic space (i.e., $P(N) \propto \frac{1}{N}$) for $1\le N \le1000$.

The resulting probability distribution for the total number of planets with civilisations is shown in Fig.\ \ref{f:civs}.  The large fraction of planet formation which has not yet taken place ($f=0.92$) implies at most an 8\% chance of us being the only civilisation the Universe will ever have.  More typically, the expected total number of planets with civilisations would be $\langle N \rangle = 12.5$.

As noted above, Earth being the first planet with a civilisation is a very conservative assumption.  For example, if the Milky Way today had another planet with a civilisation ($\sim 10^{-9}$ civilisations per habitable planet), then Earth would be at least the \textit{ten billionth} planet with a civilisation in the observable Universe.  Generalising the problem, we suppose that Earth is the $E$th planet with a civilisation. By Poisson statistics, the larger $E$ is, the better the relative constraints on $N$ will be.  This is exactly analogous to an exposure time calculation: the more photons $E$ that arrive in an 8s exposure, the better one can predict the total number of photons $N$ if the exposure were extended to a full 100s.  For planets, the generalised probability distribution is given by
\begin{equation}
\label{e:bayes2}
P(N|f,E) \propto  (N-E+1) \left(N \atop {E-1}\right) (1-f)^{E-1} f^{N-E} P(N)
\end{equation}
which is a binomial distribution with a prefactor ($N-E+1$).  As shown in Fig.\ \ref{f:civs}, the expected total number of planets with civilisations scales as $\langle N \rangle = 12.5E$, and the relative uncertainties on this total drop as $\sim \frac{1}{\sqrt{E}}$.  Hence, the more planets with civilisations which have formed before the Earth, the more likely it is for the Universe to continue forming many more in its future.

\section{Conclusions}

Current constraints on galaxy and planet formation suggest:
\begin{enumerate}
\item The Milky Way contains $\sim 10^{9}$ Earth-like and $\sim 10^{10}$ giant planets (\S \ref{s:results}).
\item The Hubble Volume contains $\sim 10^{20}$ Earth-like planets and a similar number of giant planets (\S \ref{s:results}).
\item A metallicity threshold of [Fe/H]$=-1.5$ has very limited effects on total planet counts (\S \ref{s:results}).
\item Earth-like and giant planets both formed primarily in $10^{10.5}\Msun$ galaxies; however, giant planets are much rarer than Earth-like planets in low-mass galaxies (\S \ref{s:results}).
\item Giant planets have median ages $\sim 2.5$ Gyr younger than Earth-like planets (\S \ref{s:results}).
\item The Solar System formed after 80\% of existing Earth-like planets (in both the Universe and the Milky Way), after 50\% of existing giant planets in the Milky Way, and after 70\% of existing giant planets in the Universe (\S \ref{s:form}).
\item Assuming that gas cooling and star formation continues, the Earth formed before 92\% of similar planets that the Universe will form.  This implies a $<8\%$ chance that we are the only civilisation the Universe will ever have (\S \ref{s:prob}).
\end{enumerate}

\section*{Acknowledgements}

We thank Fred Behroozi, Mario Livio, Tom Quinn, I.\ Neill Reid, Joseph Silk, Jason Tumlinson, and the anonymous referee for their very helpful comments and suggestions.  PSB was supported by a Giacconi Fellowship from the Space Telescope Science Institute, which is operated by the Association of Universities for Research in Astronomy, Incorporated, under NASA contract NAS5-26555.

Funding for SDSS-III has been provided by the Alfred P. Sloan Foundation, the Participating Institutions, the National Science Foundation, and the U.S. Department of Energy Office of Science. The SDSS-III web site is http://www.sdss3.org/.

SDSS-III is managed by the Astrophysical Research Consortium for the Participating Institutions of the SDSS-III Collaboration including the University of Arizona, the Brazilian Participation Group, Brookhaven National Laboratory, Carnegie Mellon University, University of Florida, the French Participation Group, the German Participation Group, Harvard University, the Instituto de Astrofisica de Canarias, the Michigan State/Notre Dame/JINA Participation Group, Johns Hopkins University, Lawrence Berkeley National Laboratory, Max Planck Institute for Astrophysics, Max Planck Institute for Extraterrestrial Physics, New Mexico State University, New York University, Ohio State University, Pennsylvania State University, University of Portsmouth, Princeton University, the Spanish Participation Group, University of Tokyo, University of Utah, Vanderbilt University, University of Virginia, University of Washington, and Yale University.

\appendix
\section{Additional Systematics Affecting Planet Formation and Stability}

\label{a:systematics}

Beyond galaxy star formation rates and metallicities, planet formation rates could vary with the stellar initial mass function (IMF) and the nearby stellar density.  For the stellar IMF, host star mass is known to affect both the incidence and survival time of protoplanetary disks \citep[see][for a review]{Williams11b}.  If the stellar IMF varied with redshift \citep{Dave08} or galaxy mass \citep{Weidner06}, this would change the distribution of host star masses, correspondingly changing the planet formation rate per unit stellar mass.  The Milky Way appears consistent with a non-evolving IMF \citep{Chabrier03}; more massive galaxies may have formed relatively more stars below 1 $\Msun$ (i.e., a \citealt{Salpeter55} IMF) in the past \citep{Conroy12,Sonnenfeld12}, although this has been debated \citep{Smith14,Smith15}. 

For Earth-like planets around sun-like stars, a \cite{Salpeter55} IMF yields only 30\% more mass in F5-K stars than a \cite{Chabrier03} IMF.  For giant planets, the incidence rate scales with host star mass to at least the first power for stars less than $1\Msun$ \citep{Cumming08,Kennedy08}, meaning that a \cite{Salpeter55} IMF would yield at most 70\% more giant planets than a \cite{Chabrier03} IMF.  These offsets are well within the 0.5-1 dex uncertainties on the overall planet incidence \citep{Lissauer14}.

Stellar density may impact protoplanetary disk survival either via radiation (from nearby O stars) or via interactions in dense star clusters \citep{Williams11b}.  Both would cause a redshift evolution in the planet formation rate per unit stellar mass, because galaxies at high redshifts were (on average) much denser than galaxies today \citep{vanderWel14,Shibuya15}.  For example, galaxies at $z=9$ are $\sim 20$ times smaller in every direction, resulting in galaxy densities similar to today's globular clusters \citep{Shibuya15}.  Calculations for the effects of nearby O stars suggest, however, that the impact on protoplanetary disks is limited to regions beyond 15 AU (e.g., beyond Saturn) even at the dense centres of star clusters \citep{Adams04,Clarke07}, with the effects limited to $>50$ AU (e.g., beyond Neptune) for more typical environments within clusters \citep[cf.][]{Thompson13}.  For interactions, \cite{Wang15} found that giant planet incidence in binary star systems was only reduced for binary separations less than 20 AU; incidence for binary separations between 20 and 200 AU was instead mildly enhanced, and incidence at greater separations was similar to individual stars' incidence rates.  As a comparison, stellar densities in globular clusters correspond to typical stellar separations of $\sim$ 50000 AU; even for the longest-lived (10 Myr) protoplanetary disks, cluster stars will have typical closest approaches of $\sim$ 1000 AU \citep{Adams06}.

While planet formation rates may be relatively insensitive to the nearby environment, the same cannot be said of planet orbit stability over several Gyr \citep[see][for a review]{Davies14}.  Fortuitous microlensing events have suggested significant numbers of free-floating giant planets \citep{Sumi11,Strigari12}, which could have scattered or migrated from their original birthplaces but may also have formed \textit{in situ} \citep{Veras12}.  In clusters, flybys of stars near planetary systems cause direct ejection, longer-term ($\sim$100 Myr post-flyby) destabilization, and increased orbit eccentricities \citep{Malmberg11}.  That said, observational evidence for different planet incidence rates in clusters has been mixed \citep{Sigurdsson03,Burke06,Montalto07,Quinn12,Meibom13}.

Accounting for these effects is beyond the scope of this paper.  Instead, we note that typical stars in the Milky Way are 5--10 Gyr old \citep{BWC13}, so that planets detected around these stars (e.g., with \textit{Kepler}) are exactly those that have remained bound for long periods of time.  Hence, the PFR is this paper is best interpreted as the \textit{bound} planet formation rate.  Even so, there is no guarantee that these planets remained in stable orbits for the lifetimes of their host stars; even mild changes in planet eccentricity could be detrimental to the development of civilisations (\S \ref{s:prob}).  Qualitatively, larger stellar densities in the past would lead to more interactions and therefore less planetary stability.  This would reduce the fraction of habitable planets formed before the Earth, proportionally raising the likelihood of future civilisations according to the argument presented in \S \ref{s:prob}.

\section{Recovering Galaxy Star Formation Histories}

\label{a:bwc13}

Our reconstruction technique (detailed fully in \citealt{BWC13}) uses forward modelling to extract the relationship between stellar mass, halo mass, and redshift $(M_*(M_h,z))$.  Briefly, we adopt a flexible parametrisation\footnote{The $z=0$ stellar mass---halo mass relation has six parameters: characteristic $M_*$ and $M_h$, a faint-end slope, a bright-end shape, a faint--bright transition shape, and the scatter in $M_*$ at fixed $M_h$.  For each $z=0$ parameter, another variable controls the evolution to intermediate ($z \sim 1$) redshifts, and a third variable controls the evolution to high ($z>3$) redshifts.} for $M_*(M_h,z)$.  Any $M_*(M_h,z)$ in this parameter space may be applied to a dark matter simulation, assigning galaxy stellar masses to every halo at every redshift.  Linking haloes across redshifts with merger trees, the implied evolution of galaxy stellar mass, as well as average galaxy star formation rates, can be reconstructed.  At the same time, the resulting predictions for observables, including galaxy number densities, galaxy specific star formation rates, and total cosmic star formation rates are available from the assigned stellar masses and inferred star formation rates.  Comparing these predictions to observations using a Markov Chain Monte Carlo method, we are able to constrain the allowable form of $M_*(M_h,z)$, and consequently the allowable reconstructions for galaxy star formation histories.

Observational constraints are compiled in \cite{BWC13} from over $40$ recent papers.  At low redshifts, these include results from SDSS and PRIMUS \citep{Moustakas13}; at high redshifts, these include constraints on galaxy number densities and star formation rates for $z<8$ from Hubble observations \citep{Bouwens11b,Bouwens11,McLure11,BORG12}.  Note that, above, we generate predictions for galaxy number densities as a function of stellar mass and redshift ($\phi(M_*,z)$).  For calculations involving galaxy number densities in this paper, we use the prediction for $\phi(M_*,z)$ from the best-fitting $M_*(M_h,z)$; this gives a smooth form for $\phi(M_*,z)$ which is less susceptible to sample variance than using the individual data sources in our compilation.

We use the \textit{Bolshoi} dark matter simulation \citep{Bolshoi} for halo properties (including mass functions and merger rates).  \textit{Bolshoi} follows a periodic, comoving volume 250 $h^{-1}$ Mpc on a side with 2048$^3$ particles ($\sim 8\times 10^9$), each with mass $1.9 \times 10^8$ $\Msun$, and was run with the \textsc{art} code. \citep{kravtsov_etal:97,kravtsov_klypin:99}  The adopted cosmology (flat $\Lambda$CDM; $h=0.7$, $\Omega_m = 0.27$, $\sigma_8=0.82$, $n_s = 0.95$) is consistent with WMAP9 results \citep{WMAP9}.  Halo finding and merger tree construction used the \textsc{rockstar} \citep{Rockstar} and \textsc{Consistent Trees} \citep{BehrooziTree} codes, respectively.

{\footnotesize
\bibliography{master_bib}
}


\end{document}